\documentstyle[12pt,epsf]{article}
\newcommand{\bea}{\begin{eqnarray}}
\newcommand{\eea}{\end{eqnarray}}
\newcommand{\bean}{\begin{eqnarray*}}
\newcommand{\eean}{\end{eqnarray*}}
\newcommand{\ba}{\begin{array}}
\newcommand{\ea}{\end{array}}
\newcommand{\be}{\begin{equation}}
\newcommand{\ee}{\end{equation}}
\newcommand{\nn}{\nonumber}

\newcommand{\bra}[1]{\langle #1|}
\newcommand{\ket}[1]{|#1\rangle}
\newcommand{\av}[1] {\langle #1\rangle}

\newcommand{\amp}[2]{\langle #1|#2\rangle}

\newcommand{\pa}{\partial}
\newcommand{\ga}{\gamma}
\newcommand{\ep}{\epsilon}
\newcommand{\half}{\mbox{\scriptsize $\frac{1}{2}$}}
\newcommand{\hh}{\hat H}
\newcommand{\ha}{\hat a}
\newcommand{\hAA}{\hat A}
\newcommand{\hN}{\hat N}
\newcommand{\hp}{\hat \psi}
\newcommand{\hphi}{\hat \phi}
\newcommand{\hpd}{\hat \psi ^\dagger}
\newcommand{\he}{\hat E}
\newcommand{\hb}{\hat b}
\newcommand{\hjo}{\hat j _0}

\newcommand{\leave}{\! \! \! \! \! / \, \,}
\newcommand{\lm}{\lim _{y \rightarrow x}}
\newcommand{\scd}{\partial ^2 _{A_T}}

\newcommand{\fd}[1]{\frac{\delta }{\delta #1}} 
\newcommand{\refpa}[1]{(\ref{#1})} 

\newcommand{\tr}{\mbox{${\rm tr~}$}}
\newcommand{\ra}{\rightarrow}
\newcommand{\lr}{\leftrightarrow}

\newcommand{\etat}{\eta_T}

\newcommand{\hpt}{\hat \psi_T}

\newcommand{\om}{\omega}
\newcommand{\Om}{\Omega}
\newcommand{\sgn}{\mbox{sgn}}

\newcommand{\gammat}{\tilde{\gamma}}
\newcommand{\prd}[3] {Phys.\ Rev.\ D               {#1} {(#2)} {#3}}
\newcommand{\annp}[3]{Ann.\ Phys.\ (N.Y.)          {#1} {(#2)} {#3}}
\newcommand{\pr}[3]  {Phys.\ Rev.                  {#1} {(#2)} {#3}}

\newcommand{\binom}[2] {{#1\choose #2}}

\newcommand{\hj}{\hat j}
\newcommand{\hQ}{\hat Q}
\newcommand{\hJ}{\hat J}
\newcommand{\hA}{\hat A}
\newcommand{\hH}{\hat H}
\newcommand{\de}{\delta}

\begin{document}
\begin{titlepage}
\begin{flushleft}
G\"oteborg\\
ITP 95-29\\
hep-th/9601136\\
December 1995\\
\end{flushleft}
\vspace{0.5cm}
\begin{center}
\vspace{1.5cm}

{\large \bf
Explicit gauge invariant quantization of the Schwinger model on a circle 
in the functional Schr\"odinger representation}\\
\vspace{5mm}
{  Joakim Hallin\footnote{Email address: tfejh@fy.chalmers.se}
and Per Liljenberg\footnote{Email address: tfepl@fy.chalmers.se}}\\
\vspace{1cm}
{\sl Institute of Theoretical Physics\\
Chalmers University of Technology\\
and University of G\"oteborg\\
S-412 96 G\"oteborg, Sweden}\\

\end{center}
\begin{abstract}
We solve the Schwinger model on a circle by first finding the explicit groundstate  
functional(s). Having done this, we give the structure of the Hilbert space and derive 
bosonization formulae in this formalism.
\end{abstract}
\end{titlepage}
\section{Introduction and Discussion}
Massless QED in 1+1 dimensions was first studied by Schwinger \cite{schwinger}.
 Since then it has 
been considered by numerous authors. In 
particular it has been studied on the circle by \cite{manton85,hetrick88,iso88}.

The
spectrum of the theory is equivalent to that of a massive free boson. From a fermionic 
point of view this result is non-trivial. In this paper we are treating 
the Schwinger model in a functional representation. This has some nice properties. 
We treat gauge-invariance in Dirac's way i.e. we let states be annihilated by 
the constraint operator and we only allow operators commuting with the constraint 
operator. In a functional representation this can be done explicitly. Furthermore,
the functional representation space is large enough to contain the inequivalent
Fock spaces that always exist in quantum field theory. In the case of the
Schwinger model this is desirable since the solution of the model usually 
involves a Bogoliubov transformation that takes you out of a chosen Fockspace
into another. The functional representation allows you to treat all these Fock spaces
in a general setting.

One of the main motivations for this paper is the the study of the 
groundstate functional for the Schwinger  model, 
expressed in terms of fermionic variables. The idea is
that  the groundstate in more complicated 
theories, like the massive Schwinger model or higher dimensional QED, may be
 of the same or a similar form as that of the Schwinger model. 
Even if this does not turn out 
to be the case, one might use it as a variational ansatz for such theories. 

So what does the groundstate look like? 
The groundstate cannot be gaussian in a fermionic representation since the spectrum is 
not that of a free fermion. 
It turns out that the groundstate is a two parameter 
functional where one of the parameters is the covariance associated with the 
free (gaussian) groundstate and the other parameter is induced by the interaction. 
We would like to stress that the groundstate is annihilated by Gauss' law.
This holds in particular for the free groundstate. There is therefore no need
of a modified Gauss' law as suggested in \cite{kiefer94}.

The non-gaussian character of the groundstate has some unpleasant properties. Trying 
to evaluate expectation values directly, using our general expression for the inner 
product, one runs into very hard calculations. However for numerical calculations 
our inner product is well suited. To calculate expectation values in a simple way we 
can use bosonization.

The sequence of deriving various statements of the model is 
different in this work than in previous works. We start out by finding the 
groundstate. This fixes the irreducible representation within the 
functional 
representation that we are using. Having done this, we derive the anomalous chiral 
current algebra and find the creation operators and the structure of the
Hilbert space. We also give a proof of bosonization by showing that the action of 
the fermionic operators and the corresponding bosonic ones have the same action on 
the explicit groundstate. Finally we calculate the different 2-point
correlators of the theory. 

This is a technical paper. We have tried to keep it readable by
putting all proofs and lengthy calculations in appendices.

\section{Hamiltonian}
The Hamiltonian of massless QED on a circle of circumference $L$ is,
\be \hh =\int _0 ^L \!\!dx \left\lgroup \half\he ^2 (x)-\hpd (x) \ga (i \pa _x-e \hAA (x)) \hp (x)\right\rgroup,
\ee
where $\ga =\ga ^0 \ga ^1$. In an explicit representation take $\ga ^0 =\sigma _1$ , 
$\ga ^1=-i \sigma _2$ and hence $\ga =\sigma _3$. One also has the first 
class constraint operator, Gauss' law:
\be
\hat G(x)=\pa _x \he (x)-e \half \lbrack \hpd (x),\hp (x) \rbrack =
\pa _x \he (x)- e \hjo (x) \approx 0 .\label{gauss}
\ee
Furthermore, boundary conditions needs to be specified. We choose,
\bea
\hAA (x+L) &=& \hAA (x), \\
\he (x+L) &=& \he (x), \\
\hp (x+L) &=& e^{-2 \pi i \alpha} \hp(x).
\eea
Since $\he$ is periodic, \refpa{gauss} implies that the total electric charge 
$\hat Q_0=e\int _0 ^L dx \hj _0(x) $ vanishes on physical states. Define the transverse 
fields,
\bea
\hAA _T &=& \frac{1}{L} \int _0 ^L dx \hAA (x),\\
\he _T &=& \frac{1}{L} \int _0 ^L dx \he (x),\\
\hp _T (x) &=& \exp {\left\lgroup i e \int _0 ^ x dx' \hAA (x')-i e x \hAA _T+i 2 \pi 
\alpha x/L \right\rgroup } \hp (x).
\eea
With this definition, $\hp _T$ is periodic, $\hp _T(x+L)=\hp _T (x)$. Under a  
gauge transformation $\Lambda (x)=e^{i \lambda (x)}$, the fields transform as
\bea
\he '(x) &=& \he (x),\\
\hAA '(x) &=& \hAA (x)-\frac{1}{e}\pa _x \lambda (x),\\
\hp '(x) &=& \Lambda (x) \hp (x).
\eea
Under a small gauge transformation, $\lambda (x)$ periodic and $\lambda (0)=0$, 
we find that $\hAA _T$ and $\hp _T$ are invariant. Under a global transformation 
$\lambda (x)=\lambda (0)$, $\hAA _T$ is still invariant while 
$\hp _T' =e^{i \lambda (0)} \hp _T$.
Finally, for a large transformation (picking a representive for 
each class), $\lambda (x)=2 \pi n x/L$, we have
\bea
\hAA _T ' &=& \hAA _T-\frac{2 \pi n}{e L},\\
\hp _T '(x) &=& e^{i 2 \pi n x/L} \hp _T(x).
\eea
The fields satisfy the following nonvanishing commutators and anticommutators:
\bea\label{eq:com}
\lbrack \hAA (x),\he (y) \rbrack &=& i \delta (x-y),\nn\\
\{ \hp _\alpha (x),\hpd _\beta (y) \} &=& \delta _{\alpha \beta} \delta (x-y) .
\eea
Let us also introduce the operator $\hJ$, which commutes with $\hat G$,
\be
\hJ _{\alpha \beta}(x,y) = \half \lbrack \hpd _{T,\alpha} (x),\hp _{T,\beta} 
(y) \rbrack .
\ee
Using \refpa{eq:com} we obtain,
\bea
\lbrack \hAA _T,\he _T \rbrack &=& \frac{i}{L} ,\\
\lbrack \hJ _{\alpha \beta} (x,y),\hJ _{\alpha ' \beta '} (x',y') \rbrack &=& 
\delta _{\beta \alpha '} \delta (y-x') \hJ _{\alpha \beta '} (x,y')-
\delta _{\alpha \beta '} \delta (x-y') \hJ _{\alpha ' \beta}(x',y) ,\label{eq:jcom} \\ 
\lbrack \hJ (x,y) ,\he _L(z) \rbrack &=& e\left(\frac{y-x}{L}-\int _x^y dx' 
\delta (x'-z)\right) 
\hJ (x,y) ,\\
\lbrack \hJ (x,y) ,\he _T \rbrack &=&0 .
\eea
Note that \refpa{eq:jcom} only holds on all states as long as $x \neq y$ or 
$x' \neq y'$. The constraint operator now reads,
\be
\hat G(x)=\pa _x \he _L (x)-e \hjo (x)=\pa _x \he _L (x)-e \half \lbrack 
\hpd _T(x), \hp _T(x) \rbrack =\pa _x \he _L (x)-e\hJ _{\alpha \alpha}(x,x),
\ee
where $\he _L(x)=\he (x)-\he _T$.  
Hence on the set of physical states, defined such that $\hat G(x)=0$ on that set, 
we may 
express $\he _L(x)$ in terms of $\hjo (x)=\hJ _{\alpha \alpha}(x,x)$. Doing 
this one may write the Hamiltonian on the set of physical states as,
\be
\hh =\frac{L}{2} \he _T ^2-i \int _0 ^L dx \lm (\pa _y+i e \hat{b}) \ga _{\alpha 
\beta} \hJ _{\alpha \beta}(x,y)+\int _0 ^L dx\int _0^L dy \bar V(x-y) \hat j_0(x) \hat 
j_0(y),
\ee
where $\hb=\hAA _T-\frac{2 \pi \alpha}{e L}$ and
\be
\bar V(x)=\frac{e^2L}{4\pi ^2}\sum _{n>0} \frac{1}{n^2} \cos{\frac{2\pi n x}{L}}.
\ee
\section{Functional representation}
To implement the canonical
(anti)commutation relations \refpa{eq:com}
we will use the functional Schr\"odinger representation. For fermions we will
use the reducible representation first found in \cite{jackiw88}. 

Let the bosonic operators  
$\hAA(x), \he(x)$ and the fermionic 
operators $\hp(x), \hpd(x)$ act on
wavefunctionals $\Psi (A,\eta ^* , \eta )$ of a real bosonic field $A(x)$,
a complex Grassmann field $\eta(x)$
and its complex conjugate $\eta ^*(x)$ according to
\begin{eqnarray}
\hAA(x)& \lr & A(x) ,\\
\he(x) & \lr & \frac{1}{i}\fd{A(x)} ,\\
\hp _{\alpha} (x) & \lr & \frac{1}{\sqrt{2}}(\eta _{\alpha} (x)+\fd{\eta
_{\alpha} ^*(x)}),
\label{eq:psi}\\
\hpd _{\alpha} (x) & \lr & \frac{1}{\sqrt{2}}(\eta ^* _{\alpha} (x)+\fd{\eta
_{\alpha} (x)}).
\label{eq:psid}
\end{eqnarray}
A general wave functional may be viewed as an 
overlap \cite{hallin95} with a product of a bosonic field state
 and a Grassman field state, 
$\ket{A \eta ^* \eta}=\ket{A}\otimes\ket{\eta ^* \eta}$,
\begin{eqnarray}
\Psi (A,\eta ^* , \eta ) &=& \amp{A \eta ^* \eta}{\Psi}, \nn \\
\amp{A \eta ^* \eta}{A'{\eta '} ^* \eta '} &=& \delta(A-A')
	\exp \left[\int dx \left(\eta_{\alpha}^* (x)
 	{\eta'}_{\alpha}(x)-\eta ^{'*}_{\alpha} (x) \eta _{\alpha} (x)\right)\right].
\end{eqnarray}
Here, and in the following, left out integration limits means that the integral
should be taken from $0$ to $L$.
Using the field states, the partition of unity and the inner product are 
given by functional integration \cite{jackiw88,hallin95},
\bea
\hat 1 & = & \int DA \ket{A} \bra{A}  \otimes 
	\int D^2 \eta ' D^2 \eta \ket{\eta ^* \eta}\amp{\eta ^* \eta}
	{{\eta '} ^* \eta '}\bra{{\eta '} ^* \eta '}, \\
\amp{\Psi _1}{\Psi _2} &=&\int DA \,D^2 \eta ' D^2 \eta \; \amp{\eta ^* \eta}
	{{\eta '} ^* \eta '} \; \Psi_1 ^*(A,\eta ^*,\eta) 
	\Psi _2(A,{\eta '}^*,\eta ').\label{eq:inner1}
\eea
where we have set $D^2 \eta=D \eta ^* D\eta$. 

\section{Gauge-invariant states}

Gauge-invariant states are annihilated by Gauss' law, 
\be
\hat{G}(x)\Psi(A,\eta ^* , \eta )=0.
\ee
These states are invariant under the transformations that are generated by 
$\hat{G}$, i.e. the small and the global gauge transformations. 
Gauge-invariant wavefunctionals have 
been found in \cite{hallin93}. A general 
gauge-invariant functional is parameterized by a family of 
distributions denoted
${f}$ and has the following form (spinor indices are summed over):
\bea\label{eq:gauge-inv}
\lefteqn{\Psi_{f}(A,\eta,\eta^*)=\bigg[ f^{(0)}(A_T)+} \nn\\ 
& & +\sum _{a=1}^\infty \frac{1}{a!}
	\int \! d^a\! x \, d^a\! y \, f^{(a)}(A_T;x_1,y_1,\ldots ,x_a,y_a)
	 \,\etat^*(x_1)\gammat\,\etat(y_1)\cdots
	\etat^*(x_a)\gammat\,\etat(y_a)\bigg]\nn \\ 
& & \times \exp \left[\int dx \, \etat^*(x) M \, \etat(x)\right]
,\label{eq:genst}
\eea
where the real variable $A_T$ and the Grassman field $\eta_{T,\alpha}(x)$ are
 defined in analogy with $\hAA_T$ and $\hpt(x)$, i.e.
\bea
A_T&=&\frac{1}{L} \int _0 ^L dx A (x) \\
\eta_{T,\alpha}(x)&=&\exp {\left[ i e \int _0 ^ x dx' A(x')-i e x A _T+i 2 \pi 
\alpha x/L \right] } \eta_{\alpha} (x).
\eea
A few comments are in order. The constant matrix $M$ in \refpa{eq:gauge-inv}
must satisfy
$\tr M=0$ and $M^2= 1$. These conditions do not uniquely
determine $M$. However, due to the reducibility of the representation 
different choices of $M$ give  functionals all representing the same physical
state. Furthermore, the matrix $M$ determines the matrix  $\gammat$ through
$\gammat=\frac{1}{2} (1-M)\gamma(1+M)$. A convenient choice of $M$ giving a 
gaussian free groundstate is $M=i\gamma^1$. In the following that choice is understood.
In the gauge-invariant state \refpa{eq:gauge-inv}, the parameterizing 
distributions multiply "base states",
\be\label{eq:basest}
\etat^*(x_1)\gammat\,\etat(y_1)\cdots
	\etat^*(x_a)\gammat\,\etat(y_a)
	\exp \left[\int dx \, \etat^*(x) M \, \etat(x)\right].
\ee
The base states are invariant under the exchange of pairs $(x_i,y_i) \lr (x_j,y_j)$
but change sign under exchange of  $x_i \lr x_j$ or $y_i \lr y_j$ separately.
We will demand that the distributions $f^{(a)}$ satisfy
\be\label{eq:symm}
f^{(a)}(A_T;\ldots ,x_i,y_i,\ldots ,x_j,y_j,\ldots)=
	f^{(a)}(A_T;\ldots ,x_j,y_j,\ldots ,x_i,y_i,\ldots).
\ee
It is also worth mentioning
that gauge-invariant states characterized by translation-invariant distributions
(in the spatial coordinates) have zero total momentum.

\subsection*{Action of physical operators on gauge-invariant states}

Physical (gauge-invariant) operators may be constructed from the
operator $\hJ_{\alpha \beta}(x,y)$. Let the operator 
$A\cdot \hJ(x,y) = A_{\alpha \beta} \hJ_{\alpha \beta}(x,y)$ act on
a general gauge-invariant state
parameterized by the family $f$. The state produced will be a 
gauge-invariant state parameterized by a new family, say $f_A$, i.e.
\be
A\cdot\hJ(x,y) \Psi_f = \Psi_{f_A}
\ee
The resulting families $f_A$ 
may be  expressed in terms of the original family $f$.
In \cite{hallin93} the families $f_A$ were found for
$A$ equal to $1,\gamma^0,\gamma^1,\gamma$ respectively. 
For families having the property \refpa{eq:symm} the result is
\bea
\lefteqn{
f_1^{(a)}(A_T;x_1,y_1,\ldots ,x_a,y_a)=}\nn\\
	&& \sum _{b=1}^a \left\{ \delta (x-x_b) 
	f^{(a)}(A_T;x_1,y_1,\ldots ,y,y_b,\ldots ,x_a,y_a)\right. \nn\\
	&&~~~\left.-\delta (y-y_b)
	f^{(a)}(A_T;x_1,y_1,\ldots ,x_b,x,\ldots ,x_a,y_a)\right\},
	 \label{eq:oneact}\\
\lefteqn{
f_{\gamma^0}^{(a)}(A_T;x_1,y_1,\ldots ,x_a,y_a)=}\nn\\
	&& -i f^{(a+1)}(A_T;x_1,y_1,\ldots ,x_a,y_a,y,x)\nn \\
	&& +i \sum _{b=1}^a \Big\{ 
	f^{(a+1)}(A_T;x_1,y_1,\ldots ,x_b,x,y,y_b,\ldots ,x_a,y_a)
	\nn\\
	&&~~~~~ + f^{(a-1)}(A_T;x_1,y_1,\ldots ,x_b\leave ,y_b \leave ,
	\ldots ,x_a,y_a) \delta (x-x_b)\delta (y-y_b) \Big\}, \label{eq:gnact}\\
\nn\\
\nn\\
\nn\\
\lefteqn{
	f_{\gamma^1}^{(a)}(A_T;x_1,y_1,\ldots ,x_a,y_a)=}\nn\\
	&&-i\delta (x-y) 
	f^{(a)}(A_T;x_1,y_1,\ldots ,x_a,y_a) \nn\\
	&&+i\sum _{b=1}^a \left\{ \delta (x-x_b) 
	f^{(a)}(A_T;x_1,y_1,\ldots ,y,y_b,\ldots ,x_a,y_a)\right.\nn \\
	&&\left.~~~~~~+\delta (y-y_b)
	f^{(a)}(A_T;x_1,y_1,\ldots ,x_b,x,\ldots ,x_a,y_a)\right\},
	\label{eq:goact} \\
\lefteqn{f_{\gamma}^{(a)}(A_T;x_1,y_1,\ldots ,x_a,y_a)=}\nn\\
	&&f^{(a+1)}(A_T;x_1,y_1,\ldots ,x_a,y_a,y,x)\nn \\
	&&-\sum _{b=1}^a f^{(a+1)}(A_T;x_1,y_1,\ldots ,x_b,x,y,y_b,\ldots ,x_a,y_a)\nn\\
	&&+\sum _{b=1}^a 
	f^{(a-1)}(A_T;x_1,y_1,\ldots ,x_b\leave ,y_b \leave ,\ldots ,x_a,y_a)
	\delta (x-x_b)\delta (y-y_b) .
	\label{eq:gact}
\eea
These expressions are also valid for $a=0$  
if sums ranging from one to zero are set to zero.

\subsection*{Inner product and gauge-invariant states}

Gauge-invariant state functionals are completely specified 
by a family of distributions. The inner product \refpa{eq:inner1},
when involving gauge-invariant states only, must therefore be a mapping
from a pair of families to the complex numbers. We will see how that 
comes about.

 First, considering \refpa{eq:inner1}, we observe that $DA={\rm const~} dA_T\, DA_L$
, $D^2 \eta ' D^2 \eta=D^2 \etat ' D^2 \etat$ and $\amp{\eta ^* \eta}
{{\eta '} ^* \eta '}=\amp{\etat ^* \etat}{{\etat '} ^* \etat '}$. Then,
after these changes of variables, all dependence
of $A_L$ has disappeared from the integrand and the integral over $A_L$
gives just another divergent constant. The inner product \refpa{eq:inner1}
thus becomes
\be
\amp{\Psi _g}{\Psi _f}=N\,\int_{-\infty}^{\infty} dA_T\int D^2 \etat ' D^2 \etat \;
	 \amp{\etat ^* \etat}
	{{\etat '} ^* \etat '} \; \Psi_g ^*(A_T,\etat ^*,\etat) 
	\Psi _f(A_T,{\etat '}^*,\etat ').\label{eq:inner2}
\ee
On inserting the expression \refpa{eq:gauge-inv} for the two general gauge-invariant
functionals, it is possible to do the fermionic integrals. After dropping the factor
$N$ and a factor $ \det(2 I)$ which emerges in the calculation ($I_{\alpha \beta}(x,y)=
\delta _{\alpha \beta} \delta (x-y)$) we arrive at the
final form of the inner product on the space of gauge-invariant
functionals:
\bea\label{eq:blin}
\lefteqn{\amp{\Psi _g}{\Psi _f}=\int_{-\infty}^{\infty} dA_T 
	\bigg\{{g^{(0)}}(A_T)^* f^{(0)}(A_T)}\label{eq:inner3}\\
	&&+\sum _{a=1}^\infty \frac{1}{a!} 
	\int d^a\! x \, d^a\! y \, \ep _{i_1\cdots i_a} 
	{g^{(a)}} (A_T;x_1,y_{i_1},\ldots ,x_a,y_{i_a})^* \,
	f^{(a)}(A_T;x_1,y_1,\ldots ,x_a,y_a) \bigg\}. \nn
\eea
We will return to this expression below. Another way of deriving \refpa{eq:blin} is 
by making an appropriate ansatz and then using the hermiticity properties of the 
various gauge invariant operators to constrain the ansatz as was done in 
\cite{hallin93}.

\section{Groundstates}
Having set up the formalism we are now ready to
find eigenstates of $\hh$. Using \refpa{eq:oneact} and
 \refpa{eq:gact} one finds the action of 
$\hh$ on a general gauge-invariant state $\Psi _f (A_T,\etat ,\etat ^*)$.
We have
\be
\hh \Psi _f (A_T,\etat ,\etat ^*) = \Psi _{f'} (A_T,\etat ,\etat ^*),
\ee
where the family  $f'$ is given, in terms of the family $f$, by
 (remember that $b=A_T-\frac{2\pi \alpha}{eL}$)
\bea 
\lefteqn{f'^{(a)} (A_T; x_1,y_1,\ldots ,x_a,y_a)= -\frac{1}{2L} \scd f^{(a)} (A_T; x_1,y_1,\ldots ,x_a,y_a)} \nn \\
&& -i \int dx \lm 
(\pa _y +i e b)f^{(a+1)} (A_T; x_1,y_1,\ldots ,x_a,y_a,y,x)\nn \\
&&+ i\int dx \lm (\pa _y +i e b)\sum _{b=1}^a 
f^{(a+1)} (A_T; x_1,y_1,\ldots ,x_b,x,y,y_b,\ldots ,x_a,y_a)\nn \\
&& -i \sum _{b=1}^a f^{(a-1)}(A_T;x_1,y_1,\ldots ,x_b\leave ,y_b\leave ,\ldots ,x_a,y_a) (\pa _{x_b}+i e b) \delta (x_b-y_b)\nn \\
&& +V(x_1,y_1,\ldots ,x_a,y_a) f^{(a)}(A_T; x_1,y_1,\ldots ,x_a,y_a) .\label{eq:hier}
\eea
This expression holds for $a=0$ if sums ranging from one to zero are set to zero.
The potential $V$ is defined as
\be
V(x_1,y_1,\ldots ,x_a,y_a)=\sum _{i,j=1}^a V(x_i-y_j)-\sum _{j > i=1}^a 
\left[V(x_i-x_j)+V(y_i-y_j)\right],
\ee
where
\be
V(x)=\frac{e^2L}{2 \pi ^2}\sum _{n>0}\frac{1}{n^2}\left(1-\cos{\frac{2 \pi n x}{L}}\right) =
2[\bar V(0)-\bar V(x)].
\ee
There is a simple physical interpretation of $V$. It gives rise to 
an attractive interaction between states of different charge, $(xy)$, and a repulsive 
interaction between states of equal charge, $(xx)$ or $(yy)$.

Demanding that $\Psi _f(A_T,\etat ,\etat ^*)$ is an eigenstate of $\hh$ 
with eigenvalue $E$ , i.e. 
\be
f'^{(a)}=E f^{(a)},  \, \, a=0,1,2,\ldots , \label{eq:eigens}
\ee
leads by \refpa{eq:hier} 
to a complicated set of hierarchy equations coupling different levels (we call $a$ 
the level). 

It turns out that there are eigenstates that factorize in an electromagnetic (EM) part
and a fermionic (F) part (this is not true if we add a mass 
term to the Hamiltonian), i.e. 
\bea
\Psi(A_T\eta_T^*\eta_T)&=&\Psi(A_T)\Psi(\eta_T^*\eta_T),\\
\ket{\Psi}&=&\ket{\Psi}_{EM}\otimes\ket{\Psi}_{F}.
\eea
Accordingly their families also factorize,
\be
f^{(a)}(A_T,x_1,y_1,\ldots,x_a,y_a)=\Psi (A_T) f^{(a)}(x_1,y_1,\ldots,x_a,y_a), \, \, a=0,1,2,\ldots .
\ee
For such states the strategy is: First, split the hamiltonian in an electromagnetic and
a fermionic part defined through $\hh=\frac{1}{2} \he _T^2+\hh _F$ and
find $A_T$-independent eigenstates of 
$\hh _F$ with eigenvalue $E(A_T)$.

This corresponds to solving \refpa{eq:eigens} ignoring the kinetic term for $A_T$
in the expression for $f'^{(a)}$ given in \refpa{eq:hier}.
Second, having found such eigenstates of $\hh_F$, \refpa{eq:eigens}  and \refpa{eq:hier}
collapse into
\be\label{eq:at}
-\frac{1}{2L} \scd \Psi (A_T)+E(A_T) \Psi (A_T)=E \Psi (A_T) ,
\ee
which is then solved. 

We will pursue the strategy outlined above. 
In Appendix \ref{app:eigenstates} we show that the states $\Psi _N(\eta_T^*\eta_T)$,
parameterized 
by the $A_T$-independent family $f_N$ are eigenstates 
of $\hh _F$ with eigenvalues $E_N(A_T)$. The family is given by $f^{(0)}_N=1$ and 
for $a \neq 0$,
\be
f^{(a)}_N(x_1,y_1,\ldots ,x_a,y_a)= \Om _N(x_1-y_1) \cdots \Om _N(x_a-y_a)
\Phi ^{(a)} (x_1,y_1,\ldots ,x_a,y_a),
\ee
where 
\bea
\Om _N (x) &=& -\frac{1}{L} \sum _n \sgn(n+N) e^{ip_n x},\\
\Phi ^{(a)} (x_1,y_1,\ldots ,x_a,y_a) &=& \exp{\left\lgroup \sum _{i,j=1}^a \varphi (x_i-y_j)-
\sum _{j>i=1}^a[\varphi (x_i-x_j)+\varphi (y_i-y_j)] \right\rgroup },
\eea
and
\be
\varphi(x) = -\sum _{n>0}\frac{1}{n p_n}\left(\sqrt{p_n^2+
M^2}-p_n\right) (1-\cos{p_n x}).
\ee
Here we have introduced the notation $p_n=\frac{2\pi n}{L}$ for discrete momenta and
the standard notation $M=e/\sqrt{\pi}$.
Note that when $e=0$, $f^{(a)}_N$ is just a
product of $\Om_N$'s and the corresponding functional is gaussian.
The eigenvalue $E_N(A_T)$ is, after regularization,
\be
E_N(A_T)=-\frac{\pi}{6L}+\frac{2 \pi}{L}
\left(\frac{e A_T L}{2 \pi}-N-\frac12 -\alpha \right)^2+
\sum _{n>0} \left(\sqrt{p_n^2+M^2}-p_n\right) .
\ee
From the form of the eigenenergy we see that the remaining quantum
mechanical system \refpa{eq:at} is a harmonic oscillator with well-known
solutions. Denote its lowest energy state by 
$\Psi _N(A_T)$ and its eigenvalue by $E_N$. Thus,
\bea
\Psi _N(A_T) &=& \exp\left\lgroup -\frac{2\pi}{ M L}
\left(\frac{e A_T L}{2\pi}-N-\frac12-\alpha\right)^2
\right\rgroup\\
E_N &=& -\frac{\pi}{6L}+\frac{M}{2}+\sum _{n>0} \left(\sqrt{p_n^2+M^2}-p_n\right).
\eea
Hence the states
$\Psi _N(A_T\eta_T^*\eta_T)=\Psi _N(A_T)\Psi _N(\eta_T^*\eta_T)$,
 parameterized by the 
family \newline $\Psi _N (A_T) f_N$, are eigenstates of $\hh$,
 all with the same energy $E=E_N$. These states 
are the groundstates of $\hh$. The corresponding  
kets will be denoted $\ket{\Psi_N}$.

The groundstates $\ket{\Psi_N}$ have an important property.
Under a large gauge transformation $\lambda (x)=2\pi x/L$, we have that
\bea
A_T' &=& A_T-\frac{2 \pi}{eL},\\
\etat '(x) &=& e^{i 2\pi x/L}\etat (x),
\eea
which implies that $\ket{\Psi _N}$ is mapped into $\ket{\Psi _{N+1}}$.
 Hence a state transforming only by a phase 
under large gauge transformations is the $\theta$-vacuum $\ket{\theta}$ defined as,
\be
\ket{\theta}=\sum _Ne^{-iN\theta} \ket{\Psi _N}.
\ee

\section{Currents, charges  and creation operators}

In the last section we found the explicit groundstate(s) of the model.
 The entire Hilbert
space may now be constructed by the action of physical operators on the groundstate(s).
The different groundstates $\ket{\Psi_N}$ all define 
different representation spaces of the
algebra of the physical operators. These representation spaces need not
be orthogonal. However, using a set of well-known creation operators
one may construct orthogonal Fock spaces from the different groundstates.
Then, an additional operator connects the different Fock spaces.

In this section we will define some operators and investigate their properties
within the functional representation. These operators will be needed
when calculating expectation values in the next section.

We start by defining the currents and their fourier transforms. Let
\bea\label{eq:currdef}
\hj_0(x)&=&1 \cdot \hJ (x,x),\\
\hj_5(x)&=&\ga \cdot \hJ (x,x),\\
\hj_\pm (x)& =& \half\left(\hj_0(x)\pm\hj_5(x)\right)=
 \frac{1}{L} \sum_{n} \hj_\pm(n) e^{i p_n x}.
\eea 
Hermiticity demands that 
$\hj_\pm^\dagger(n)=\hj_\pm(-n)$. When acting on the 
representation space(s) defined
by the groundstate(s) $\ket{\Psi_N}$,
the chiral currents have a well-known anomalous commutator, a Schwinger term,
\be\label{eq:anomaly1}
[\hj_\pm (x),\hj_\pm (y)]=\pm \frac{1}{2\pi i} \de'(x-y).
\ee
In Appendix \ref{anomaly} we verify this algebra in the functional representation
by calculating its action on the explicit groundstate $\Psi_N(A_T\eta_T^*\eta_T)$.
In momentumspace the algebra reads
\be\label{eq:anomaly2}
[\hj_\pm (n),\hj_\pm^\dagger (m)] = \pm n\delta_{n,m}.
\ee
The chiral charge is defined as,
\be
\hQ _5=\int dx \hj _5 (x) .
\ee
Let us act with $\hQ_5$ on a groundstate. By \refpa{eq:gact} we obtain
\be
\hQ_5 \Psi_N(A_T\eta_T^*\eta_T) =  L\Om_N(0)\;\Psi_N(A_T\eta_T^*\eta_T).
\ee 
The expression for $\Om_N(0)$ contains a sum which is not absolutely convergent.
To make the sum well-defined we use for consistency the exponential regularization
used in appendix \ref{app:eigenstates}. We find
\be
\hQ_5 \Psi_N(A_T\eta_T^*\eta_T) =  
2 (\frac{eA_T L}{2\pi}-N-\frac12-\alpha)\;\Psi_N(A_T\eta_T^*\eta_T).
\ee 
From the above expression we deduce the following commutators, valid when
acting on the representation spaces defined by the groundstates:
\bea
[\hQ_5,A\cdot\hJ(x,y)]&=&[\ga,A]\cdot\hJ(x,y),\\
\lbrack\hQ_5,\he_T]&=&i\frac{e}{\pi}. \label{eq:anom}
\eea
In particular note the anomalous commutator \refpa{eq:anom}.
On the  representation spaces defined
by the groundstates $\ket{\Psi_N}$ we may thus write
\be
\hQ_5 = 2 (\frac{e\hA_T L}{2\pi}-\hN-\frac12-\alpha),
\ee 
where the operator $\hN$ 
is defined through its action on the groundstate(s),
\be
\hN \ket{\Psi_N}_{EM} = \ket{\Psi_N}_{EM}~,~\hN \ket{\Psi_N}_F =N \ket{\Psi_N}_F,
\ee
and through its commutator with physical operators, 
\bea\label{eq:ncom}
&[\hN,A\cdot\hJ(x,y)]=[-\half\hQ_5,A\cdot\hJ(x,y)]=\half[ A,\ga ]\cdot\hJ(x,y)&,\\
&[\hN,\hA_T]=[\hN,\he_T]=0.&
\eea
Note that the regularization 
has made $\hQ_5$ invariant under large gauge-transformations. Furthermore,
the groundstate 
$\ket{\Psi_N}$ is not an eigenstate of $\hQ_5$. 

One can find creation operators $\ha_\pm^\dagger(n)$, which
are related to the currents through
a Bogoliubov transform. We have
\be
\left[\ba{c} \ha_+^\dagger(n) \\ \ha_-(n) \ea \right] =
 \half \left[ \ba{cc}  \kappa _n+\kappa _n^{-1} & \kappa _n-\kappa _n^{-1}\\ \kappa _n-\kappa _n^{-1} & \kappa _n+\kappa _n^{-1}
\ea \right] \left[\ba{c} \hj_+^\dagger(n) \\ \hj_-^\dagger(n) \ea \right]~,~n\neq 0,
\ee
where $\kappa _n=\left[1+\frac{M^2}{p_n^2}\right]^{\frac{1}{4}}$ 
and as before $\ha_\pm^\dagger(n)=\ha_\pm(-n)$.
When acting on the representation spaces defined by the groundstates, 
the algebra \refpa{eq:anomaly2}
leads to 
\bea
[\ha_\pm(n),\ha_\pm^\dagger(m)]&=&n \delta_{n,m},\\
\lbrack \hat{H} , \ha_\pm^\dagger(n)]&=&\sqrt{p_n^2+M^2}~ \ha_\pm^\dagger(n)~,~n>0.
\eea
Also, one may check that the operators $\ha_\pm(n),n>0$
annihilate $\ket{\Psi_N}$ and that $\ha_+^\dagger(n),n>0$ ($\ha_-^\dagger(n),n>0$) 
create states with
positive (negative) momenta.

Apart from the creation operators above there
is the creation operator related to the electromagnetic sector creating states 
with zero momentum,
\be
\ha^\dagger=\sqrt{\frac{L}{2M}}
\left(M\hA_T-i\he_T -\frac{2 \pi M}{e L}(N+\half+\alpha)\right)=\sqrt{\frac{L}{2M}}
\left(\frac{\pi M}{eL}\hQ _5-i\he_T\right) .
\ee
Furthermore, $\ha$ annihilates $\ket{\Psi_N}$ and
\bea
[a,a^\dagger]&=& 1,\\
\,[\hH ,a^\dagger] &=& M a^\dagger.
\eea
The operator $a^\dagger$  together with the operators $\ha_\pm^\dagger(n)$ are
the creation operators building up the infinite set of orthogonal
Fock spaces from the different
groundstates. Since all the creation operators commute with $\hN$ a 
Fock space consisting of states transforming with a phase under large
gauge-transformations may be built upon the theta vacuum. 
The spectrum is that of a free massive boson.
In the following section we will
calculate expectation values and further examine the concept of bosonization
 in the context of the functional representation.

\section{Expectation values and bosonization}
Let us first discuss some general properties of overlaps. 
States with different eigenvalues of $\hat N$ 
are orthogonal since $\hat N$ is hermitian. By calculating the $N$-charge of various 
operators one may then figure out between what states these operators have 
non-vanishing expectation values. From \refpa{eq:ncom} we have e.g.
\bea
\lbrack \hat N ,\half(\ga \pm 1) \cdot \hat J(x,y)\rbrack &=& 0,\\
\lbrack \hat N, \half(\ga ^0 \pm \ga ^1) \cdot \hat J(x,y) \rbrack &=& \pm 
\half(\ga ^0 \pm \ga ^1) \cdot \hat J(x,y) .
\eea
Consider thus e.g.,
\be\label{eq:toeval}
\frac{ \bra{\Psi _N} \half(\ga ^0-\ga ^1) \cdot \hat J(x,y) \ket{\Psi _{N+1}}}
{\parallel \ket{\Psi _N} \parallel \; \parallel \ket{\Psi _{N+1}}\parallel} =
\frac{ \bra{\Psi _N} \half(\ga ^0-\ga ^1) \cdot \hat J(x,y) \ket{\Psi _{N+1}}}
{\amp{\Psi _N}{\Psi _N}}.
\ee
Since $\ket{\Psi _N}=\ket{\Psi _N}_{EM}\otimes \ket{\Psi _N}_F$ is a product state 
we can evaluate the fermionic and 
electromagnetic part separately. Trying to use the inner product 
\refpa{eq:inner3} to evaluate 
\refpa{eq:toeval} directly, one runs into very hard calculations. This is
due to the non-gaussian character of $\ket{\Psi _N}_F$.
The fermionic vacuum amplitude $ _F\amp {\Psi _N}{\Psi _N}_F$ (which 
happens to be a very complicated object) doesn't 
factorize from the numerator of \refpa{eq:toeval} in a straightforward manner.
However, to show that \refpa{eq:inner3} works in principle, 
we have evaluated \refpa{eq:toeval} 
to first order in $\varphi (x)$. One finds for the fermionic part:
\bea
\lefteqn{ _F\bra{\Psi _N}\half(\ga ^0-\ga ^1)
 \cdot \hat J(x,y) \ket{\Psi _{N+1} }_F=}\nn\\
&& =\frac{2i}{L} \sum _{a=0}^\infty \left[\binom{\lambda -1}{a}
+\sum _{n>0} \left\{\binom {\lambda -1}{a} C_n (1-\cos{p_n (x-y)}) 
+\binom{\lambda -3}{a} 4 C_n (n-1)\right\} \right] \nn\\
&& =\frac{i}{L} 2^\lambda \left(1+\sum _{n>0} nC_n -C_n \cos{p_n (x-y)}\right),
\eea
\be
 _F\amp {\Psi _N}{\Psi _N}_F
= \sum _{a=0}^\infty 
\left[\binom{\lambda }{a}+\binom{\lambda -2}{a} \sum _{n >0} 4n C_n\right] =
2^\lambda \left(1+\sum _{n>0} n C_n\right)
\ee
where $\lambda =L \delta (0)=\sum _n$ is assumed to be regularized in a suitable 
manner and where $C_n=(1-\sqrt{1+M^2/p_n^2})/n$. Hence the fermionic part of 
\refpa{eq:toeval} to first order in $C_n$ becomes,
\be\label{eq:jlex2}
\frac{ _F \bra{\Psi _N} \half(\ga ^0-\ga ^1) \cdot \hat J(x,y) \ket{\Psi _{N+1}}_F}
{ _F\amp{\Psi _N}{\Psi _N}_F}=
\frac{i}{L} \left(1-\sum _{n>0} C_n \cos{p_n (x-y)}\right) .
\ee
In order to evaluate \refpa{eq:toeval} exactly, and to verify \refpa{eq:jlex2}, 
we will use bosonization. 
In appendix \ref{app:bosonization} we show the following formulae by calculating
their action on the explicit groundstate in the functional representation:
\bea\label{eq:bos0-1}
\half(\ga ^0-\ga ^1) \cdot \hat J(x,y)& =&\frac{i}{L} \hat S 
e^{\frac{2\pi i \hat N}{L}(x-y)} e^{\hphi_+^\dagger(x)-\hphi_-^\dagger(y)}
e^{-\hphi_+(x)+\hphi_-(y)}, \\
\half(\ga ^0+\ga ^1) \cdot \hat J(x,y)& =&-\frac{i}{L} 
e^{\frac{2\pi i \hat N}{L}(x-y)} e^{\hphi_-^\dagger(x)-\hphi_+^\dagger(y)}
e^{-\hphi_-(x)+\hphi_+(y)}\hat{S}^\dagger, \\
(\ga \pm 1) \cdot \hat J(x,y) &=& \Om _0(y-x) e^{\frac{2\pi i \hat N}{L}(x-y)}
e^{\hphi_\pm^\dagger(x)-\hphi_\pm^\dagger(y)}e^{-\hphi_\pm(x)+\hphi_\pm(y)}.
\eea
Here we have defined
$\hphi_\pm(x)=\sum _{n>0} \frac{1}{n} \hj_\pm(\pm n) e^{\pm ip_n x}$.
The shift operator $\hat S$ shifts the fermionic part of $\ket{\Psi _N}$ into 
the fermionic part of $\ket{\Psi _{N-1}}$,
\bea
\hat S \ket{\Psi _N}_F &=&\ket{\Psi _{N-1}}_F,\\
\hat S^\dagger \ket{\Psi _N}_F &=&\ket{\Psi _{N+1}} _F,
\eea
and leaves the electromagnetic part
invariant, i.e. the state 
$(\hat S \Psi _N)(A_T,\etat ,\etat ^*)$ is 
parameterized by the family $\Psi _N(A_T) f_{N-1}$.
Having the bosonization formulae  it is a simple task
 to calculate vacuum expectation values. One 
expresses the currents in terms of annihilators and creators and normal orders 
the exponentials using $e^A e^B=e^{A+B} e^{\half[A,B]}$.

In view of what we will eventually evaluate, namely $\theta$ expectation values, define 
the operator $\hat J_l$ invariant under large gauge transformations,
\be\label{eq:jl}
\hat J_l(x,y) =e^{-i e \hat A_T (x-y)} \hat J(x,y) .
\ee
We obtain,
\bea\label{eq:jlex1}
\frac{ \bra{\Psi _{N'}} \half(\ga ^0\pm \ga ^1) \cdot \hat J_l(x,y) \ket{\Psi _{N}}}
{\amp{\Psi _N}{\Psi _N}}&=&\mp\frac{i}{L} \delta _{N',N\pm 1}e^{-\frac{\pi}{ML}-\frac{2\pi i\alpha }{L} 
(x-y)-\frac{\pi M}{4L} (x-y)^2} g_1(x-y), \\
\frac{ \bra{\Psi _{N'}} \half(\ga \pm 1) \cdot \hat J_l(x,y) \ket{\Psi _{N}}}
{\amp{\Psi _N}{\Psi _N}}&=&\delta _{N',N}\Om _0(y-x) 
e^{-\frac{2\pi i}{L}(\alpha+\half)(x-y)-\frac{\pi M}{4L} (x-y)^2} g_2(x-y) ,\; \; \; \; 
\; \; 
\eea
where
\bea
g_1(x)&=&\exp \left\lgroup\sum _{n>0} \frac{1}{n}\left(1 -\half(\kappa _n^2+\kappa _n^{-2})
+\half(\kappa _n^2-\kappa _n^{-2}) \cos{p_n x}\right)\right\rgroup ,\\
g_2(x)&=& \exp \left\lgroup\sum _{n>0} \frac{1}{n} \left(1-\half(\kappa _n^2+\kappa _n^{-2})\right)
(1-\cos{p_n x})\right\rgroup .
\eea
We also recall that $\kappa _n=\left[1+\frac{M^2}{p_n^2}\right]^{\frac{1}{4}}$.
By expanding $g_1(x)$ to first order in $C_n=\frac{1}{n}(1-\kappa _n^2)$
 it is easy to see that the fermionic 
part of \refpa{eq:jlex1} and \refpa{eq:jlex2} agrees. Furthermore when $x=y$ 
\refpa{eq:jlex1} agrees with the result found in \cite{hetrick88} up to an 
irrelevant phase which can 
be absorbed in the definition of $\ket{\Psi _N}$. Finally defining $\theta$ 
expectation values by $\av{A}_\theta=\bra{\theta}A\ket{\theta}/\amp{\theta}{\theta}$ 
one obtains
\bea\label{eq:thex}
\av{\half(\ga ^0\pm \ga ^1) \cdot \hat J_l(x,y)}_\theta &=&\mp \frac{i}{L} 
e^{\pm i \theta-\frac{\pi}{ML}-\frac{2\pi i\alpha }{L} 
(x-y)-\frac{\pi M}{4L} (x-y)^2} g_1(x-y), \\
\av{1 \cdot \hat J_l(x,y)}_\theta &=&0,\\
\av{\gamma \cdot J_l(x,y)}_\theta &=& \Om _0(y-x) 
e^{-\frac{2\pi i}{L}(\alpha+\half)(x-y)}e^{-\frac{\pi M}{4L} (x-y)^2} g_2(x-y) .
\eea
To conclude this paper we use \refpa{eq:thex} to evaluate the chiral condensate,
\be
\av{\hat{\bar \psi} (x) \hat \psi (x)}_\theta =-2 \sin \theta e^{-\frac{\pi}{ML}} g_1(0)
\rightarrow -\frac{M}{2\pi} e^\ga \sin \theta , \; \; L\rightarrow \infty .
\ee
\newpage 
\appendix
\section{Eigenstates}\label{app:eigenstates}
We want to find eigenstates of $\hh _F$, the fermionic part of the Hamiltonian. Let 
these states be parameterized by the $A_T$-independent family $f_N$ where 
$N$ is an integer. Denoting the eigenenergy by $E_N(A_T)$ we should solve
\be
\label{eq:tosolve}\hh _F \Psi _{f_N}(\etat,\etat ^*)=
E_N(A_T) \Psi _{f_N}(\etat,\etat ^*),
\ee
or in terms of the family $f_N$,
\be\label{eq:tosolve2}
f_N^{'(a)}=E_N(A_T) f_N^{(a)}
\ee
where $f_N^{'(a)}$ is calculated using \refpa{eq:hier} ignoring the kinetic term
for $A_T$. Make the ansatz $f_N^{(0)}=1$ and for $a \neq 0$,
\be
f_N^{(a)}(x_1,y_1,\ldots ,x_a,y_a)=\Phi ^{(a)}(x_1,y_1,\ldots ,x_a,y_a) 
\Om _N(x_1-y_1) \cdots \Om _N(x_a-y_a),
\ee
where
\bea
\label{eq:omn}\Om _N(x) &=& -\frac{1}{L}\sum _n \sgn (n+N) e^{i p_n x} ,\\
\Phi ^{(a)} (x_1,y_1,\ldots ,x_a,y_a) &=& 
\exp{\lbrack \sum _{i,j=1}^a \varphi (x_i-y_j)-
\sum _{j>i=1}^a(\varphi (x_i-x_j)+\varphi (y_i-y_j)) \rbrack },
\eea
and 
\be
\varphi (x)=\sum _{n>0}C_n(1-\cos{p_n x}) .
\ee
Furthermore we have defined,
\be
\sgn (n)=\left\{ \begin{array}{ll} 1, & n\geq 0 \\ -1, &  n < 0  \end{array}\right. .
\ee
For some purposes it is convenient to rewrite $\Om _N$ in the following manner,
\be
\Om _N(x)=\delta (x)+\om _N(x),\label{eq:conv}
\ee
where
\be
\om _N(x)=-\frac{2}{L} \sum _{n\geq -N} e^{i p_n x}=-\frac{2}{L} e^{- i p_N x} \sum _{n\geq 0} e^{ i p_n x} =e^{- i p_N x} \om _0(x).
\ee
Some important properties of $\Om _N$ are established by Fourier transforming the 
product $g(x) \Om _N(x)$ for any (periodic) function $g(x)$:
\bea
\label{eq:lim1}\lim _{x \rightarrow 0} (g(x) \Om _N(x)) &=& \frac{1}{i\pi} g'(0)+g(0) \Om _N(0),\\
\label{eq:lim2}\lim _{x \rightarrow 0} (g(x) \Om _N'(x)) &=& \frac{i}{2\pi} g''(0)+g(0) \Om _N'(0) .
\eea
The singular quantities $\Om _N(0)$ and $\Om _N'(0)$ have to be regularized in a 
suitable manner for them to make sense. When analyzing $\refpa{eq:hier}$ we will 
also use the following property of $\Phi$,
\bea \label{eq:phi}
\lefteqn{\Phi ^{(a+1)}(x_1,y_1,\ldots ,x_a,y_a,y,x) =}\nn\\
&&=\Phi ^{(a+1)}(x_1,y_1,\ldots ,x_b,x,y,y_b, 
\ldots ,x_a,y_a) \nn \\
&& =\Phi ^{(a)}(x_1,y_1,\ldots ,x_a,y_a) \nn \\
&& \times \exp{\lbrack \varphi (y-x)+\sum _{i=1}^a 
\varphi (x-x_i)-\varphi (y-x_i)+\varphi (y-y_i)-\varphi (x-y_i) \rbrack} .
\eea

The eigenvalue $E_N(A_T)$ is easy to find. By \refpa{eq:tosolve2} we have that
$E_N(A_T)=f'^{(0)}_N$. Remembering to ignore the kinetic term, \refpa{eq:hier}
leads to
\be
E_N(A_T)=-i\int dx \lm (\pa _y+i e b)f_N^{(1)}(y,x)=-\frac{L}{2\pi}\varphi ''(0)
+L(eb\Om_N(0)-i \Om_N'(0)),\label{eq:en}
\ee
having used $\refpa{eq:lim1}$ and $\refpa{eq:lim2}$. We will regularize the singular 
part of $E_N$ by exponential regularization. Hence write,
\bea
L(eb\Om_N(0)-i \Om_N'(0))_\ep &=& -\frac{2\pi}{L}\sum _n(n+\frac{ebL}{2\pi}) \sgn(n+N) 
e^{-\ep |n+\frac{ebL}{2\pi}|}\nn \\
&=& -\frac{2\pi}{L}\left[\frac{2}{\ep ^2}+\frac{1}{12}-\left(\frac{ebL}{2\pi}
-N-\frac12 \right)^2 \right]+{\cal{O}}(\ep) .
\eea
Subtracting the pole in $\ep$ and letting $\ep \rightarrow 0$ the expression for 
the regularized energy is,
\be
E_N(A_T)=-\frac{L}{2\pi}\varphi ''(0)-\frac{\pi}{6L}+\frac{2 \pi}{L}
\left(\frac{e A_T L}{2 \pi}-N-\frac12-\alpha \right)^2 .
\ee
Now let's find the eigenstates. Using \refpa{eq:hier} and \refpa{eq:en} the 
equation \refpa{eq:tosolve2} with  $a=1$ reduces to
\bea
\lefteqn{-i(\pa _{x_1}+ieb)\delta (x_1-y_1)+W(x_1-y_1) \Om _N(x_1-y_1)}\nn \\
&& +i\int dx\Om _N(x_1-x)(\pa _x+ieb+\varphi '(x-y_1)
-\varphi '(x-x_1))\Om _N(x-x_1)=0,\label{eq:level1}
\eea
where $W(x)=V(x)+D(x)$ and
\[
D(x_1-y_1)=-\frac{1}{2\pi}\int dx(\varphi '(x-y_1)-
\varphi '(x-x_1))^2=-\sum _{n>0} C_n^2 n p_n (1-\cos{p_n (x_1-y_1)}) .
\]
Similarly one finds for $a=2$, using \refpa{eq:level1},
\bea\label{eq:level2}
\lefteqn{0=i \int dx \bigg\lgroup \Om_N(x_1-y_1) 
\lbrack \delta (x_2-y_2) \delta (x_2-x)-\Om_N(x_2-x) \Om_N(x-y_2) \rbrack} \\
&&~~~~~~~~~~~~~\times (\varphi '(x-x_1)-\varphi '(x-y_1)) 
+ (1 \leftrightarrow 2)\bigg\rgroup\nn\\
&&+\Om_N(x_1-y_1) \Om_N(x_2-y_2)
[ W(x_1-y_2)+W(x_2-y_1)-W(x_1-x_2)-W(y_1-y_2)].\nn
\eea
If \refpa{eq:level1} and \refpa{eq:level2} are satisfied then \refpa{eq:tosolve2} 
is satisfied for all $a$. Using \refpa{eq:conv} one may obtain the identity,
\bea \label{eq:g}
\lefteqn{i\int dx\varphi '(x-x_1) \Om _N(x_2-x) \Om _N(x-y_2)}\nn \\
&& =i \delta (x_2-y_2) \varphi '(x_2-
x_1)+\Om _N(x_2-y_2) (G(x_2-x_1)-G(y_2-x_1)),
\eea
where,
\be
G(x)= \sum _{n>0}C_n p_n (1-\cos{p_n x}) .
\ee
Hence \refpa{eq:level2} becomes,
\bea
\lefteqn{0= \Om _N(x_1-y_1) \Om _N(x_2-y_2) \bigg\lgroup W(x_1-y_2)+
2G(x_1-y_2)+W(x_2-y_1)}\nn \\
&& +2G(x_2-y_1)-W(x_1-x_2)-2G(x_1-x_2)-W(y_1-y_2)-2G(y_1-y_2)\bigg\rgroup,
\eea
i.e. $W(x)+2G(x)=0$ which leads to a quadratic equation for $C_n$. Keeping only the 
root giving a convergent series for $\varphi$ we end up with,
\be
C_n=\frac{1}{n}(1-\sqrt{1+\frac{e^2L^2}{4n^2\pi ^3}})=-\frac{1}{n p_n}
(\sqrt{p_n^2+e^2/\pi}-p_n).
\ee
By \refpa{eq:omn} one has,
\be
-i(\pa _{x_1}+ieb)\delta (x_1-y_1)+i\int dx\Om _N(x_1-x)(\pa _x+ieb)\Om _N(x-x_1)=0 ,
\ee
thus reducing \refpa{eq:level1} to $W(x_1-y_1)+2 G(x_1-y_1)=0$ having used \refpa{eq:g}.
\section{Anomalous commutators}
\label{anomaly}
We will verify the chiral current algebra on the
groundstate. First regularize the 
current operators by point splitting
\be
\hj_\pm (x) = A_\pm \cdot \hJ (x,x) ~\ra~ \hj_\pm (x,\ep) = A_\pm \cdot \hJ (x,x+\ep)
\ee
where $A_\pm=\half(1 \pm \gamma)$. From \refpa{eq:jcom} it then follows that
\be\label{eq:com2}
[\hj_\pm (x,\ep),\hj_\pm (y,\ep)]=\de(x-y+\ep) A_\pm \cdot \hJ (x,y+\ep)
	-\de(x-y-\ep) A_\pm \cdot \hJ (y,x+\ep)
\ee
Now act with the commutator on the  groundstate $\Psi_N(A_T,\eta_T^*,\eta_T)$
 parameterized by
the family $\Psi_N(A_T) f_N$. Let the resulting state  be
parameterized by a family $\Psi_N(A_T) f'_\pm$, i.e.
\be
\lim_{\ep \ra 0}\;\; [\hj_\pm (x,\ep),\hj_\pm (y,\ep)]\;\Psi_N(A_T,\etat^*,\etat)
	=\Psi_N(A_T)\Psi_{f'_\pm}(\etat^*,\etat)
\ee
By \refpa{eq:oneact},\refpa{eq:gact} and \refpa{eq:com2} we obtain
\bea
f'^{(0)}_\pm&=& \pm \half \lim_{\ep\ra 0} \left\{ \de(x-y+\ep) f_N^{(1)}(y+\ep,x)-
	 \de(x-y-\ep) f_N^{(1)}(x+\ep,y)\right\} \nn\\
	&=& \pm \half \lim_{\ep\ra 0} \left[ \{ \de(x-y+\ep)- \de(x-y-\ep)\}
	\Om(2\ep) e^{\varphi(2 \ep)}\right] \nn\\
	&=& \pm \frac{1}{2\pi i} 
	\frac{\pa}{\pa (2\ep)}\left[\{\de(x+y+\ep)
	-\de(x+y-\ep) \}e^{\varphi(2 \ep)}\right]_{\ep=0}\nn\\
	&=&\pm \frac{1}{2\pi i} \de'(x-y) f_N^{(0)}
\eea
Here we have used the property \refpa{eq:lim1} of $\Om_N$ and that $\varphi'(0)=0$. 
For the higher levels we simply have
\bea
\lefteqn{f'^{(a)}_\pm(x_1,y_1,\ldots,x_a,y_a) =} \nn\\
&=& \pm \half \lim_{\ep\ra 0} \left\{ \de(x-y+\ep)
	 f_N^{(a+1)}(x_1,y_1,\ldots,x_a,y_a,y+\ep,x) \right.\nn\\
	 &&\left.~~~~~~~~-\de(x-y-\ep) f_N^{(a+1)}(x_1,y_1,\ldots,x_a,y_a,x+\ep,y)
	\right\}.
\eea
All other terms in \refpa{eq:oneact} and \refpa{eq:gact} 
are regular as $\ep \ra 0$ and automatically cancel. 
A similar calculation to the one above yields
\be
f'^{(a)}_\pm(x_1,y_1,\ldots,x_a,y_a)=
\pm \frac{1}{2\pi i} \de'(x-y) f_N^{(a)}(x_1,y_1,\ldots,x_a,y_a)
\ee
and therefore we have
\be
[\hj_\pm (x),\hj_\pm (y)]\;\Psi_N(A_T,\etat^*,\etat)=
	\pm \frac{1}{2\pi i} \de'(x-y)\,\Psi_N(A_T,\etat^*,\etat) 
\ee
Since the commutator is a $c$-number, this result holds not only for the
groundstate but for the entire
representation space defined by the groundstate(s).

\section{Bosonization}\label{app:bosonization}
To prove the bosonization formulas, we simply show that the fermionic and bosonic 
operators have the same action on the groundstate $\ket{\Psi _N}$. 
Acting with the different fermionic operators on $\ket{\Psi _N}$ produce  states 
parameterized by 
the family $\Psi _N(A_T) f'$.  
The action of $\half(\ga ^0\pm\ga ^1) \cdot \hat J(x,y)$ gives
\be\label{eq:fermionf0}
f'^{(0)} =-\frac{i}{2}(\Om _N(y-x) \pm \delta (y-x)) e^{\varphi (y-x)},
\ee
\bea\label{eq:fermionf}
\lefteqn{f'^{(a)}(x_1,y_1,\ldots ,x_a,y_a) = 
\Phi ^{(a+1)}(x_1,y_1,\ldots ,x_a,y_a,y,x)} \nn \\
&& \times \bigg\lgroup -\frac{i}{2} (\Om _N(y-x) 
\pm \delta (y-x)) \prod _{i=1}^a \Om _N(x_i-y_i) \nn \\
&& +\frac{i}{2}\sum _{i=1}^a (\Om _N(x_i-x)\pm \delta (x_i-x)) (\Om _N(y-y_i)\pm 
\delta (y-y_i)) \prod _{j\neq i} \Om _N(x_j-y_j) \bigg\rgroup ,
\eea
and for $\half(\ga \pm 1) \cdot \hat J(x,y)$ we obtain
\be
f'^{(0)}=\half\Om _N(y-x) e^{\varphi (y-x)},
\ee
\bea
\lefteqn{f'^{(a)}(x_1,y_1,\ldots ,x_a,y_a) = \Phi ^{(a+1)}(x_1,y_1,\ldots ,x_a,y_a,y,x)} \nn \\
&& \times \lbrack \half \Om _N(y-x) \prod _{i=1}^a \Om _N(x_i-y_i) \nn \\
&& -\half \sum _{i=1}^a (\Om _n(x_i-x) \mp \delta (x_i-x))(\Om _N(y-y_i)\pm 
\delta (y-y_i)) \prod _{j\neq i} \Om _N(x_j-y_j) \rbrack .
\eea
In the following, let $z$ denote the quantity $z=e^{i \frac{2\pi}{L}}$. We have,
\bea
\Om _N(x)+\delta (x) &=&z^{-Nx} \frac{2}{L} \sum _{n>0} z^{-nx}=z^{-Nx} \frac{2}{L} 
\frac{z^{-x}}{1-z^{-x}} ,\\
\Om _N(x)-\delta (x)=\om _N(x) &=&-z^{-Nx} \frac{2}{L}\sum _{n \geq 0} z^{nx}=
-z^{-Nx}\frac{2}{L} \frac{1}{1-z^{x}} \label{eq:form2}.
\eea
To avoid the pole in $z=1$ one may regularize by e.g. 
\[ \sum _{n \geq 0} z^{nx} e^{-\ep n}. \] We will assume that such a regularization 
has been done in otherwise ill defined expressions. Also define,
\be
\xi (x)=\sum _{n \geq 0} \frac{1}{n} z^{nx}=-\log{(1-z^x)} .
\ee
The identity $e^A e^B=e^{A+B} e^{\half \lbrack A,B \rbrack }$ valid when  
$\lbrack A,B \rbrack $ is a c-number will be used 
repeatedly. 
We will prove now prove that
\be\label{eq:form}
\half(\ga ^0-\ga ^1) \cdot \hat J(x,y)=\frac{i}{L} \hat S 
e^{\frac{2\pi i \hat N}{L}(x-y)} e^{\hphi_+^\dagger(x)-\hphi_-^\dagger(y)}
e^{-\hphi_+(x)+\hphi_-(y)},
\ee
which may be rewritten to yield
\be
\half (\ga ^0-\ga ^1) \cdot \hat J(x,y)=\frac{i}{L} \hat S z^{\hat N (x-y)} 
e^{\xi (0)} e^A,\label{eq:bos}
\ee
where
\bea
A &=&\half\int dx' \big\lgroup [\xi (x'-x)-\xi (y-x')-\xi (x-x')+\xi (x'-y)]
 \hat j_0(x ')\nn \\
&& ~~~~~~~~~+[\xi (x'-x)+\xi (y-x')-\xi (x-x')-\xi (x'-y)]\hat j_5(x') \big\rgroup .
\eea
The action of $A$ produces a state which can be written in the form $B \ket {\Psi _N}$ 
having defined $B$ as,
\be
B=\int dx' [\xi (x'-x)-\xi (y-x')+\varphi (x-x')-\varphi (y-x')] \hat j_0(x') .
\ee
Thus we have,
\be
e^{A} \ket {\Psi _N}=e^{A} e^{-A+B} \ket {\Psi _N}=e^{\half \lbrack A,B 
\rbrack} e^B \ket {\Psi _N},
\ee
where the commutator is $\half\lbrack A,B \rbrack =\xi (y-x)-\xi (0)+\varphi (y-x)$.
Hence the action of the right hand side of \refpa{eq:bos} on $\ket{\Psi _N}$ 
by \refpa{eq:oneact} gives 
a state parameterized by a family $\Psi(\hA_T)f''$, where
\be\label{eq:bosonf0}
f''^{(0)} =\frac{i}{L} z^{N(x-y)} e^{\xi (y-x)+\varphi (y-x)}
\ee
\bea\label{eq:bosonf}
\lefteqn{f''^{(a)}(x_1,y_1,\ldots , x_a,y_a)=\frac{i}{L} z^{N(x-y)} e^{\xi (y-x)+\varphi (y-x)} 
\Phi ^{(a)} (x_1,y_1,\ldots ,x_a,y_a) }\nn \\
&& \times  \prod _{i=1}^a \Om _{N-1}(x_i-y_i)
\; \; \mbox{exp} \lbrack \xi (x_i-x)-\xi (y-x_i)-\xi (y_i-x)+\xi (y-y_i)\nn\\
&& +\varphi (x-x_i)
-\varphi (y-x_i)+\varphi (y-y_i)-\varphi (x-y_i) \rbrack
\eea
Clearly, by \refpa{eq:form2} and $e^{\xi (x)}=\frac{1}{1-z^x}$, 
\refpa{eq:bosonf0} and \refpa{eq:fermionf0} coincide. For $a \neq 0$ consider,
\bea
&& \Om _{N-1}(x_i-y_i)
\; \; \mbox{exp} \lbrack \xi (x_i-x)-\xi (y-x_i)-\xi (y_i-x)+\xi (y-y_i) \rbrack \nn\\
&& =\delta (x_i-y_i) +\om _N(x_i-y_i) z^{x_i-y_i} \frac{(1-z^{y-x_i})(1-z^{y_i-x})}
{(1-z^{x_i-x})(1-z^{y-y_i})} \nn\\
&& =\delta (x_i-y_i) +\om _N(x_i-y_i) \frac{(1-z^{x_i-x})(1-z^{y-y_i})-(1-z^{x_i-y_i}) 
(1-z^{y-x})}{(1-z^{x_i-x})(1-z^{y-y_i})}\nn\\
&& =\Om _N(x_i-y_i)+\frac{L}{2} z^{-N(x_i-y_i)} (1-z^{y-x}) 
\om _0 (x_i-x) \om _0 (y-y_i) .
\eea
Now since the base states \refpa{eq:basest} change sign under the transformation $x_i 
\leftrightarrow x_j$ or $y_i \leftrightarrow y_j$ all terms in $f''^{(a)}$ invariant 
under this transformation will vanish when contracted with a base state. Dropping 
such symmetric terms and using \refpa{eq:phi} one sees that \refpa{eq:bosonf} and 
\refpa{eq:fermionf} coincide. For the other bosonic operators similar calculations 
lead to the same action as the fermionic operators on the groundstate. To prove 
equivalence for all states in the same representation space as the groundstate 
one also has to check the commutator algebra of the bosonic operators with the 
fermionic ones. We will 
leave this calculation out as it is well known.

\end{document}